\documentclass[draft]{agujournal}

\journalname{Geophysical Research Letters, \textbf{DOI:} 10.1002/2017GL075251}

\begin{document}

%
%

\title{Hierarchical Bayesian modeling of fluid-induced seismicity }

%
%

 \authors{M. Broccardo\affil{1}\thanks{Current address, HIT building F43, Wolfgang Pauli Str. 27 Z\"urich CH-8093 Switzerland},
A. Mignan\affil{1,2,3}, S. Wiemer\affil{1,2,3},
 B. Stojadinovic\affil{1,4}, and D. Giardini\affil{1,2,3}
 }
\affiliation{1}{Swiss Competence Center for Energy Research---Supply of Electricity, Zurich, Switzerland.}
\affiliation{2}{Swiss Federal Institute of Technology Zurich, Institute of Geophysics, Switzerland.}
\affiliation{3}{Swiss Seismological Service, Zurich, Switzerland.}
\affiliation{4}{Institute of Structural Engineering (IBK), Swiss Federal Institute of Technology (ETH) Zurich.}
%
%
%
%
%
\begin{keypoints}
\item{\textbf{DOI:} 10.1002/2017GL075251}

\item Bayesian framework built for classifying, analyzing, and forecasting uncertainties related to fluid-induced seismicity 
\item Robust classification and treatment of epistemic and aleatory uncertainties in a non-homogeneous Poisson framework 
\item Short-term forecast model accurately predicts the number and maximum magnitude of events
\end{keypoints}
%
%
%
\begin{abstract}
In this study, we present a Bayesian hierarchical framework to model fluid-induced seismicity. The framework is based on a non-homogeneous Poisson process (NHPP) with a fluid-induced seismicity rate proportional to the rate of injected fluid. The fluid-induced seismicity rate model depends upon a set of physically meaningful parameters, and has been validated for six fluid-induced case studies. In line with the vision of hierarchical Bayesian modeling, the rate parameters are considered as random variables. We develop both the Bayesian inference and updating rules, which are used to develop a probabilistic forecasting model. We tested the Basel 2006 fluid-induced seismic case study to prove that the hierarchical Bayesian model offers a suitable framework to coherently encode both epistemic uncertainty and aleatory variability. Moreover, it provides a robust and consistent short-term seismic forecasting model suitable for online risk quantification and mitigation.

\end{abstract}

%
%

%



%
%
%

\section{Introduction}
\label{sec:intro}

Statistical models for fluid-induced seismic events have received considerable attention in recent years as they play a key role in assessing seismic hazard \citep[see, e.g.][]{ellsworth_13, mignan_15}. Typically, fluid-induced seismicity is characterised by a time-varying seismicity rate related to the fluid injection rate {\citep[see, e.g.][]{Shap_10, dinske_shap_13, mignan_16,   mignan_17, elst_16, Lag_16, Lag_17}}. 
%
%
Although the seismicity rate changes over time, the inter-arrival times between seismic events have been shown to be statistically independent \citep{Lag_11}. In this case the non-homogeneous Poisson process (NHPP) is an ideal probabilistic model for predicting seismicity. 
However, most of current NHPP models for induced seismicity adopt a frequentist statistical approach, in which the seismicity rate, albeit unknown, is assumed to be deterministic \citep[see, e.g.][]{dinske_shap_13, bachmann_11, mena_13, mignan_17}. It is thus inferred that the uncertainties governing the problem are only aleatory (i.e. irreducible). This prevents the possibility of consistently encoding epistemic uncertainties usually related to past information arising from different sites and projects, and/or to expert judgement and beliefs (unless these are modelled using logic trees \citep{mignan_15}). Furthermore, seismicity-rate models are merely fitted to existing datasets. Although this provides a meaningful statistical description of the past events, it does not lead to a robust online forecasting model. In addition, the knowledge gained cannot be consistently encoded for future project planning. Probabilistic models based on Bayesian statistics have also been used and promoted in an induced seismicity context \citep[see, e.g.][]{wang_15, baker_16, wang_16, baker_17}. However, these studies focus mainly on the detection of changes in local tectonic seismicity rates caused by waste-fluid injection in Oklahoma. What is more, they do not explicitly apply either a statistical or physical model relating the seismicity rate to the rate of fluid injection. Therefore, defining a coherent general framework for classifying, analyzing, and forecasting uncertainties in deep fluid injections   constitutes a major step forward towards understanding and managing the risks associated with fluid-induced seismicity. 

In this study, we fulfill our research brief by presenting a hierarchical Bayesian framework. Hierarchical Bayesian models allow clear distinctions to be drawn between the sources of uncertainties as well as a consistent online updating strategy. We describe the time-varying rate of the Poissonian process as a function of the rate of fluid injection and a set of physical parameters describing underground properties. First, we apply the rate model to six fluid-induced seismicity sequences; then we transform the hyperparameters into random variables to model the uncertainties arising from different sites and statistical estimates. A major strength of the Bayesian approach is that it enables uncertainties and expert judgements about the model's parameters to be encoded into a joint prior distribution. Moreover, once the project is under way  and physical information becomes available, the Bayesian framework enables the computation of posterior distribution for the model's parameters, the formulation of predictive models for the Poissonian process and a robust forecasting strategy. 
Although we demonstrate that the proposed rate model fairly accurately describes the selected datasets, different models (e.g. based on geomechanical principles \citep{gischig_13, catalli_16, goertz_12}) or an ensemble of different models \citep{kiraly_2016} can be used without altering the structure of the proposed framework.

To explore both the benefits and the potential of the proposed framework we structured the study as follow: Section 2 introduces the NHPP process and the fluid-induced seismicity rate model; Section 3 introduces the Bayesian hierarchical model and the 'fitting' procedure; Section 4 validates the proposed {rate} model; Section 5 presents the online updating strategy and the forecasting model by testing the induced seismicity sequence  of the Basel 2006 Enhance Geothermal System (EGS) \citep{haring_08, kraft_14}. Finally, Section 6 presents a series of concluding remarks.
%
%
%
%
%
\section{Probabilistic and rate model}\label{sec:prob_model}
The recurrence of fluid-induced seismic events is characterized using an NHPP model {\citep{Shap_10, Lag_16, Lag_17, mignan_17}, }, 
\begin{linenomath*}
\begin{equation}
P(N(t)=n) = \frac{\Lambda(t;\boldsymbol\theta)^n\exp(\Lambda(t);\boldsymbol\theta)}{n!},
\end{equation}
\end{linenomath*}
where $\Lambda(t;\boldsymbol\theta)=\int_0^t\lambda(t';\boldsymbol \theta)dt'$, $\lambda(t;\boldsymbol \theta)$ is the time-varying rate of seismic events, and $\boldsymbol \theta$ is a set of model parameters. We describe $\lambda(t;\boldsymbol \theta)$ using the following piecewise function
\begin{linenomath*}
\begin{equation}
   \lambda(t;\boldsymbol \theta) = 
\begin{cases}
    10^{a_{fb}-bm_0}\dot V(t),&  t\leq t_{s},\\        10^{a_{fb}-bm_0}\dot V(t_{s})\exp\left(-\frac{t-t_{s}}{\tau}\right),         & t>t_s, \label{eq:model}
\end{cases}
\end{equation}
\end{linenomath*}
where $\dot V(t_s)$ is the injection flow rate; $\boldsymbol\theta = [a_{fb},b,\tau]$ is the set of model parameters respectively describing activation feedback, the earthquake size ratio {(i.e., the $b$ value of the Gutenberg-Richter distribution)}, and mean relaxation time; $m_0$ is the magnitude of completeness, and $t_s$ the shut-in time. 
In \eqref{eq:model}, we distinguish between the injection phase and the post-injection phase. The injection phase admits only positive fluid injection rates, and it is characterised by a linear relationship between $\dot V(t_s)$ and $\lambda(t)$ (in line with  
{\citet{Shap_10, dinske_shap_13, Hajati_15, elst_16, mignan_16, mignan_17})}. 
{Observe that $a_{fb}$ is equivalent to the Seismogenic Index in the poro-elastic context \citep{Shap_10, dinske_shap_13}. However $a_{fb}$ may also be explained by geometrical operations in a static overpressured field \citet{mignan_16}. As such, we use $a_{fb}$ as a generic statistical parameter with no preference for any underlying physical model \citep{mignan_17}.} 
The post-injection phase identifies the phase with constant null flow rate, after the injection has been terminated, and is characterised by an exponential decay typical of a diffusion process \citep{mignan_15b, mignan_16, mignan_17}. 
{Although the Modified Omori Law is sometimes used to describe post-injection seismicity \citep{Lag_10, barth_13}, \citet{mignan_17}  have shown  that an exponential function performs better than a power law for the six datasets presented in the Supplementary Material.} 

The main feature of \eqref{eq:model} is that it is fully characterised by the fluid injection profile and three physically meaningful parameters. The model was fitted with the maximum likelihood estimate (MLE) method (see Section~\ref{ba_inf}) and was validated (see Section~\ref{sec:val}) with reference to six fluid-induced seismic sequences \citep{haring_08, kraft_14, jost_1998, petty_2013, cladouhos_15, holland_13, ake_05}. 
In the Supporting Information, Table~S1 and Text~S1 provide the description and sources of the datasets, and Table~S2 shows the MLE estimates, $\boldsymbol{\hat\theta}_{MLE}$, which show the following parameter ranges: $0.77\le b\le 1.6$, $-2.4\le a_{fb}\le 0.1$, and $0.02\le\tau\le 13.7$[days]. 
%
%
%
%
%
\section{Bayesian Hierarchical model for induced seismicity}\label{sec:BHM}
The observed parameter ranges are wide, reflecting the high variability arising from different injections at various sites. We define these uncertainties as source-to-source variability. In principle, we can reduce source-to-source variability by making in-situ observations (e.g. after conducting a seismic monitoring campaign). However, data are often unavailable at the planning stage.
It follows that when planning projects we must account for this variability and later review and update it when data become available, either from an explanatory campaign or once the project has started. 
When prior information from different sources is available and model updating based on new local data is desirable, the Bayesian hierarchical approach is a viable and powerful tool. In addition, this framework allows for the inclusion of experts' opinions and judgements.

In a Bayesian approach, we consider $\boldsymbol\theta$ as a random vector,  {$\boldsymbol\Theta=[A_{fb},B,\mathcal{T}]$}, adding an extra layer of uncertainty. Parameter distributions aim to reflect the relative likelihood of possible outcomes, taking account of both source-to-source variability and the statistical uncertainties arising from parameter estimation.
To highlight the different framework, we change the notation from $\lambda(t;\boldsymbol \theta)$ to $\lambda(t|\boldsymbol \theta)$. We use $f'_{\boldsymbol\Theta}(\boldsymbol \theta)$ to denote joint prior probability distribution, which reflects our state of knowledge about the parameters $\boldsymbol\Theta$ before new in-situ observations are available.
%
%
\subsection{Prior distributions}\label{p_distr}
There are various options for selecting a prior, (e.g. see \citet{murphy_12}); in this study, we choose a subjective prior distribution, since the available data are limited to few past events, and we did not have in-situ information before the projects took place. 
In addition, we can encode the experts' judgments regarding the physical range of the parameters. We define  $f'_{\boldsymbol\Theta}(a_{fb},b,\tau)= f'_{A_{fb}}(a_{fb})f'_B(b)f'_{\mathcal{T}}(\tau)$; even though we have assumed independence in this definition, we show in Section \ref{ba_inf} that the posterior distribution encodes any type of correlation structures emerging from the data.
In this study, we select $f'_{A_{fb}}(a_{fb})=\mathcal{B}(a_{fb};p_a,q_a,l_a,u_a)$, $f'_B(b)=\mathcal{B}(b;p_b,q_b,l_b,u_b)$, where $\mathcal{B}(\cdot;\cdot)$ is the beta distribution with $p_a,q_a,p_b,q_b$ being the shape hyperparameters and $l_a, l_b,u_a,u_b$ being the lower and upper intervals of the parameters range. For $\tau$ we choose $f_{\mathcal{T}}'(\tau)=\Gamma(\tau;\alpha,\beta)$, where $\Gamma(\cdot;\cdot)$ is the gamma distribution with  $\alpha$ and $\beta$ being the shape parameters. We suggest to fix the \ {hyper}parameter $l_a, l_b,u_a,u_b$ based on experts' opinion and physical principles, while the  {hyper}parameters $p_a,q_a,p_b,q_b,\alpha,\beta$ are selected to fit MLE estimates of the six datasets used in this study. Figure \ref{fig:posteriors} shows both the marginal and the joint prior distributions for the model parameters.
%
%
\subsection{Aleatory vs epistemic uncertainties}\label{aAoE_un}
The proposed Bayesian hierarchical model enables a precise classification of uncertainties. Specifically, we follow the classical paradigm of separating epistemic uncertainties ---reducible by gathering more data or refining our models--- from aleatory uncertainties ---irreducible since they are inherently present in the model \citep{der_2009}. 

It is the modeller's duty to determine which uncertainties can and cannot be reduced. Philosophically speaking, the process of induced seismicity is a pure geomechanical problem. In principle, if we know the exact physical model and precise values of physical model parameters, the problem of predicting the occurrence and magnitude of a seismic event is deterministic. However, the physics and both model-related and statistical uncertainties are so complex that a fully deterministic prediction is not possible (analogous to natural seismicity). In this context, probabilistic modeling offers a viable language for describing both the complexity and uncertainties governing the problem. More specifically, by selecting  a Poisson process to describe the occurrence of the seismic events we implicitly assume that given a seismicity rate (either constant or time varying) no further reduction of inter-arrival time uncertainties is possible. However, when the seismicity rate model is itself a random variable (with uncertainties depending on source-to-source variability, statistical uncertainties, etc.), we assume that these uncertainties can be reduced either by gathering new data or by refining the model. So we represent the epistemic uncertainties with the joint distribution $f'_{\boldsymbol\Theta}(\boldsymbol\theta)$ and encode aleatory uncertainties in the definition of $\lambda(t|\boldsymbol\theta)$. 

In the context of Poissonian problems, classifying both these two uncertainties is not quixotic. In fact, the NHPP is a renewable process that, by the way it is constructed, only permits renewable uncertainties. By definition, aleatory uncertainties are renewable since they are immutable in time and  {ergodic}, whereas epistemic uncertainties change over time when additional information becomes available {, and, therefore, they are not-ergodic.} It follows that particular caution is called for when estimating statistics via NHPP if both these uncertainties feature in the probabilistic model. 
%
%
\subsection{Bayesian inference for non-homogeneous Poisson process}\label{ba_inf}
Given a set of observations $\mathcal{D}=[t_1,...,t_n,...,t_N; m_1,...,m_n,...,m_N]$, where $t_n$ is an occurrence time and $m_n$ a magnitude event, we update the probability distribution of the hyperparameters as follows:
\begin{equation}
	f''_{\boldsymbol\Theta}(\boldsymbol\theta|\mathcal{D})=c\mathcal{L}(\mathcal{D}|\boldsymbol \theta)f'_{\boldsymbol\Theta}(\boldsymbol \theta),
\end{equation} 
where $f''_{\boldsymbol\Theta}(\boldsymbol\theta|\mathcal{D})$ is the posterior distribution, $\mathcal{L}(\mathcal{D}|\boldsymbol \theta)$ the likelihood function, and $c=\int_{\boldsymbol\theta} \mathcal{L}(\mathcal{D}|\boldsymbol\theta)f'_{\boldsymbol\Theta}(\boldsymbol\theta)d\boldsymbol\theta$ is a normalizing factor. The posterior distribution $f''_{\boldsymbol\Theta}(\boldsymbol\theta|\mathcal{D})$ conveys our updated state of knowledge up to the time $t_N$. Once $f''_{\boldsymbol\Theta}(\boldsymbol\theta|\mathcal{D})$ is obtained, we can make predictions regarding future events, using the NHPP model based on updated uncertainties. The vehicle used for doing this is the total probability theorem, and the predictive model can be written as follows:
\begin{linenomath*}
\begin{equation}
P(N(t)=n)=\int_{\boldsymbol\theta}\left[\frac{\Lambda(t|\boldsymbol\theta)^n}{n!}\exp\left[-\Lambda(t|\boldsymbol\theta)\right]\right]f''_{\boldsymbol\Theta}(\boldsymbol\theta|\mathcal{D})d\boldsymbol\theta. \label{eq:predictive}
\end{equation}
\end{linenomath*}
The discussion of Section \ref{aAoE_un} calls for caution when formulating predictive equations in the presence of epistemic uncertainties. In fact, these  { are shared by all seismic  events, and, therefore, they} introduce dependence among the inter-arrival times.  {Consequently,  in this setting, the earthquakes cannot constitute Poissonian events \cite{der_2009}}. To face this problem, note that in \eqref{eq:predictive}, first we use the predictive NHPP \textit{conditional} for the epistemic uncertainties encoded in $\boldsymbol\theta$; then we apply the total probability theorem. 
Conversely, if first we compute $\int_{\boldsymbol\theta}\Lambda(t|\boldsymbol\theta)f''_{\boldsymbol\Theta}(\boldsymbol\theta|\mathcal{D})d\boldsymbol\theta$ and then use the NHPP, we  {operate the so-called ergodic approximation, \citep{der_2009, der_2005}.}

When the integral \eqref{eq:predictive} is computationally expensive,  instead of the full distribution $f''_{\boldsymbol\Theta}(\boldsymbol\theta|\mathcal{D})$,  {one can use the so-named plug-in approximation \cite{murphy_12}}, i.e. 
\begin{equation}
P(N(t)=n)=\int_{\boldsymbol\theta}\left[\frac{\Lambda(t|\boldsymbol\theta)^n}{n!}\exp\left[-\Lambda(t|\boldsymbol\theta)\right]\right]\delta(\boldsymbol\theta-\boldsymbol\theta^*|\mathcal{D})d\boldsymbol\theta, \label{eq:predictive2}
\end{equation}
where $\delta({\boldsymbol x})$ is the multidimensional delta Dirac function and $\boldsymbol\theta^*$ is a fixed value of the parameters. Common choices for $\boldsymbol\theta^*$ are $\boldsymbol{\bar{\theta}}$ and/or $\boldsymbol{ \hat {\theta}}_{map}$, which are the posterior mean, and the posterior mode (map in Bayesian jargon stands for maximum a posteriori estimation). 
 {
Moreover, observe that this approximation is equivalent to the frequentist approach if we use $\boldsymbol\theta_{mle}$. Despite its simplicity, this approximation under-represent our uncertainties when formulating predictions.  Moreover, it is different from the ergodic approximation, which is the result of a Bayesian average of the rate model. In this study, based on these considerations and given the simplicity of our model, we focus on the exact Bayesian prediction given by \eqref{eq:predictive}.}

 {Given the magnitude frequency distribution, $f_M(m_n|b)$, and following \cite{ogata_s}}, the likelihood function of each observation pair magnitude $m_n$ and time $t_n$ is proportional to $f_M(m_n|b)\lambda(t_n|\boldsymbol\theta)/\Lambda(T|\boldsymbol\theta)$,  {since $m_n$ and $t_n$ are statistically independent}.  {Moreover,} the probability of observing $N$ points is proportional to $\Lambda^N(T|\boldsymbol\theta)\exp(-\Lambda(T|\boldsymbol\theta))$; it follows that 
\begin{linenomath*}
\begin{equation}
\begin{split}
	\mathcal{L}(\mathcal{D}|\boldsymbol \theta) &=  \left[\prod_{n=1}^N\frac{\lambda(t_n|\boldsymbol \theta)}{\Lambda(T|\boldsymbol\theta)}f_M(m_n|b)\right]\Lambda^N(T|\boldsymbol\theta)\exp\left[-\Lambda(T|\boldsymbol\theta)\right],\\
	&=\left[\prod_{n=1}^{N}\lambda(t_n|\boldsymbol\theta)f_M(m_n|b)\right]\exp[-\Lambda(T|\boldsymbol\theta)].\label{eq:likelihood}
	\end{split}
\end{equation} 
\end{linenomath*}
The complete log-likelihood used to compute $\hat{\boldsymbol\theta}_{MLE}$ and/or $\hat{\boldsymbol\theta}_{MPA}$ is reported in the Appendix \ref{sec:loglik}.
Figure~\ref{fig:posteriors} shows the posterior distributions for the Basel 2006 case study, whereas other datasets are reported in the supplementary material. Specifically, Figure~\ref{fig:posteriors} shows, in the diagonal, the parameters' prior and posterior marginal distributions; in the lower triangular part, prior pair-wise distributions; and in the upper triangular part, posterior pair-wise distributions. As anticipated in Section~\ref{p_distr}, even though joint prior distribution is defined based on the independence of model parameters, joint posterior distribution captures the correlation structure of the problem. In particular, Figure~\ref{fig:posteriors} shows a strong correlation between activation feedback and earthquake size ratio.
 \begin{figure}[bh!]
\centering 
\includegraphics[scale=.41]{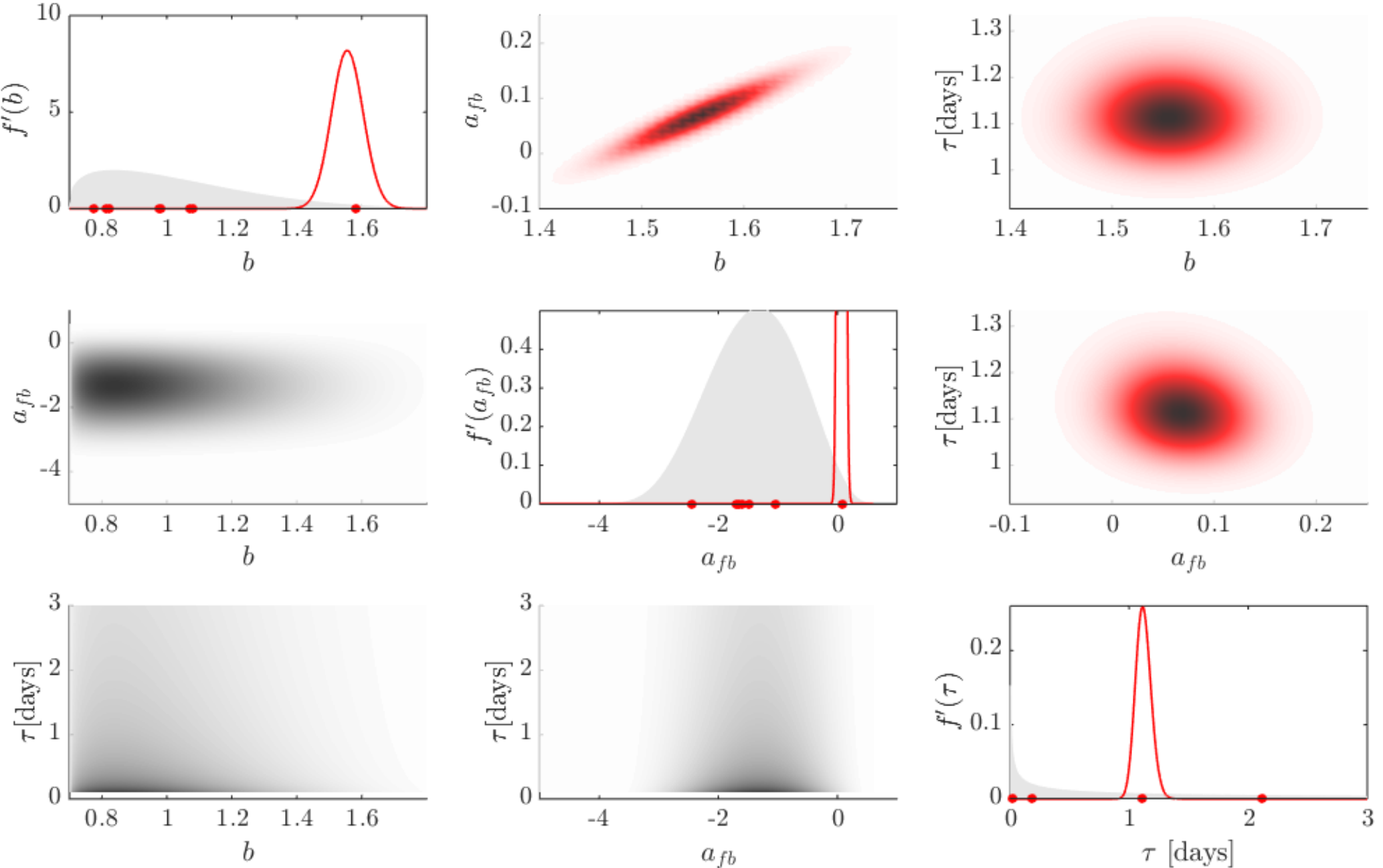}
\caption{Prior and posterior distribution for Basel 2006 dataset: $m_0$ $=$0.8, time $[0, 12]$ days magnitude $M_W$. Diagonal: in shaded grey the prior distributions, with red lines the posterior distributions, red dots represent the six $\hat{\boldsymbol \theta}_{MLE}$. Lower triangular part: joint prior distributions. Upper triangular part: joint posterior distributions.}
\label{fig:posteriors}
\end{figure}
%
%
%
%
%
 {\section{Rate model validation}\label{sec:val}}
We validate the rate model \eqref{eq:model} via the goodness of fit procedure used  by \cite{ogata_s} to verify aftershock models in a NHPP setting for tectonic seismicity.  We start by converting the dataset $\mathcal{D}^{(t)}$ into a transformed dataset as follow 
\begin{linenomath*}
\begin{equation}
\mathcal{D}^{(t)}=[t_1, ...,t_n,...,t_N]\rightarrow \tilde{\mathcal{D}}^{(\tau)}=\left[\tau_1,...,\tau_n, ...,\tau_N\right],
\end{equation}
\end{linenomath*}
where $\tau_n = \int_0^{t_n}\lambda(t;\boldsymbol\theta)dt$.
Observe that as long as the time events $t_n$ are distinct, the transformation is an isomorphism (i.e., one-to-one). Therefore, the two datasets are equivalent; however,  $\tilde{\mathcal{D}}^{(\tau)}$ has the distribution of a uniform Poisson process with unit rate. It follows that if the empirical cumulative distribution function (CDF) of $\tilde{\mathcal{D}}^{(\tau)}$, $F_{\tilde{\mathcal{D}}^{(\tau)}}(\tau)$, deviates significantly from the CDF of a uniform distribution, $F_U(\tau)$, then the model does not represent the point process properly. 

\begin{figure}[ht!]
\centering 
\includegraphics[scale=.45]{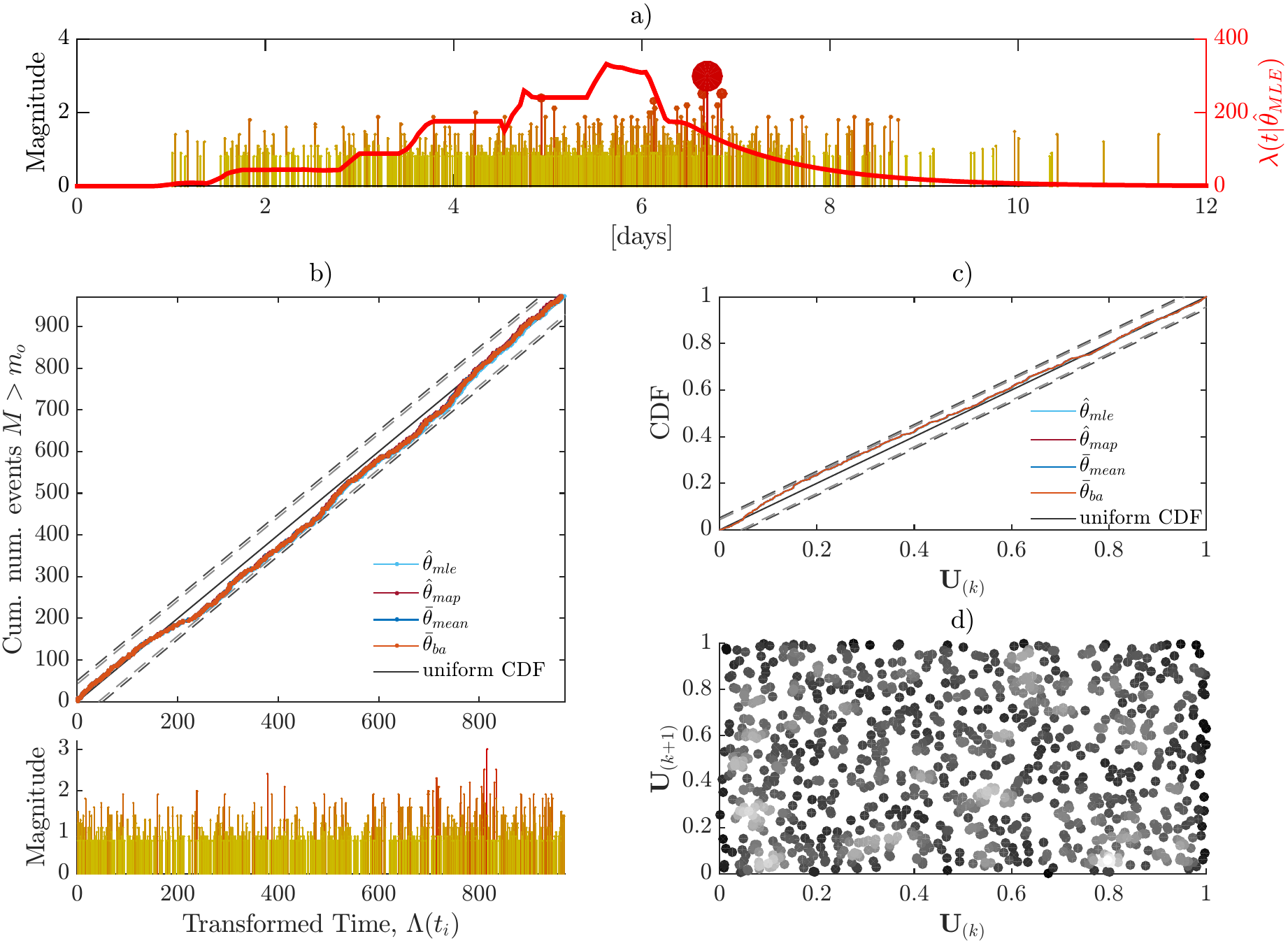}
\caption{Goodness of fit. a) Basel 2006, time series of fluid-induced time events vs magnitude, and $\lambda(t|\hat{\boldsymbol\theta}_{MLE})$. b) Cumulative number of events vs the $95\%$ and $99\%$ KS confidence intervals. Lines are overlapping. c) CDF of transformed time vs the $95\%$ and $99\%$ KS confidence intervals. Lines are overlapping. d) Test of inter-arrival independence.}
\label{fig:K-S_test_1}
\end{figure}

To verify whether the empirical CDF fits the hypothesized uniform CDF, we use the Kolmogorov-Smirnov statistic, namely $D_n = \sup_{\tau}|F_U(\tau)-F_{\tilde{\mathcal{D}}^{(\tau)}}(\tau)|$, to derive confidence intervals. We test four different rate models arising from four different selection of $\boldsymbol\theta$. In specific, we test $\Lambda(t;\hat{\boldsymbol\theta}_{mle})$, $\Lambda(t;\hat{\boldsymbol\theta}_{map})$, $\Lambda(t;\bar{\boldsymbol\theta}_{mean})$, and $\Lambda_{ba}(t)$, which are respectively the MLE estimator, the MAP estimator, the mean of the posterior distribution, and the Bayesian average of the rate model defined as $\Lambda_{ba}(t)=\int_{\boldsymbol\theta}\Lambda(t|\boldsymbol\theta)f''_{\boldsymbol\Theta}(\boldsymbol\theta |\mathcal{D})d\boldsymbol\theta$. Figure \ref{fig:K-S_test_1}a) shows the estimated $\Lambda(t;\hat{\boldsymbol\theta})$ and the seismic sequence of Basel 2006. 
Figure \ref{fig:K-S_test_1}b) shows the cumulative number of points $\tau_n$  for the four rate models versus the theoretical uniform cumulative number of points ---$NF_U(\tau)$--- for the Basel 2006 case study.  
%
%
The dashed lines represent the two-sided $95\%$ and $99\%$ Kolmogorov-Smirnov (KS) confidence intervals. For this dataset, we observe that all four models are within both confidence intervals, so in all four cases we do not have statistical evidence against the proposed model. 
In the Text~S2 and Figures S1 to S5 of the Supplementary Information we report the fit for the other five datasets.

\citet{berman} proposes an alternative test. Specifically, he uses the transformed inter-arrival times, $Y_n = \tau_n-\tau_{n-1}$, to verify whether they are independent and identically distributed (iid) exponential random variables with unit mean. Testing whether $Y_n$ are exponential iid random variables is equivalent to test whether $U_n = 1-\exp(-Y_n)$ are uniform random variables on $[0,1]$. Figure \ref{fig:K-S_test_1} c), shows the empirical CDF of $U_n$ ---$F_{U_n|\tilde{\mathcal{D}}^{(\tau)}}(u)$--- for the four rate models versus the theoretical CDF ---$F_U(u)$. As in the first test, we observe that all four models are within both confidence intervals. In addition to this analysis, Berman proposes verifying the independence of intra-arrivals. Specifically, he suggests plotting $U_{(n+1)}$ versus $U_n$. If there is any intra-arrival correlation, it should emerge in the proposed plot in the form of a cluster of points. Figure~\ref{fig:K-S_test_1} d) shows this analysis for $\Lambda(t;\hat\theta_{mle})$ model. We obtained similar graphs for the other rate models (which we do not report, to avoid redundancy). For this dataset we find no evidence against the independence of intra-arrival times. Given these analyses, we conclude that there is no evidence in the data against the NHPP model and the proposed fluid-induced rate model. To the best of our knowledge, these statistical tests have not been proposed yet to validate fluid-induced seismicity rate models, and they should be promoted to test both novel rate models and the Poissonian framework.

\section{Online updating and earthquake forecast model}
Although the outlined inference procedure is a powerful tool, it requires the full set of data before deriving the posterior distribution. Therefore, it is suitable only for the statistical analysis of past events. In this section, we tackle the more compelling problems of online parameter updating and short-term forecasting of seismic events. Building on the Basel 2006 case study, we show that the present Bayesian hierarchical model allows not only a rapid online updating procedure to reduce epistemic uncertainties, but also a reliable short forecast of the number and magnitude of incoming fluid-induced seismic events. We assume that we know the scheduled injection profile. However, in this case we do not know in advance if, and when, there a shut-in event will occur. So the likelihood differs from \eqref{eq:likelihood}, depending on whether the injection has been permanently stopped. To distinguish between the two likelihood functions, we refer to  the likelihood used in the online updating strategy as a ``partial likelihood function''.

 \subsection{Partial likelihood for on-line updating}
 Given the rate model \eqref{eq:model},  an injection scenario $\dot V(t)$, and a set of partial observations $\mathcal{D}(t)=[t_1,...,t_{\underline N}; m_1,...,m_{\underline N}]$ with $t_1<...<t_{\underline N}\le t\le t_s$, where $t_{\underline N}$ is the time up to the $\underline N$ event and $t_s$ is a future, but yet unknown shut-in time, the partial likelihood is reduced to 
 \begin{equation}
	\mathcal{L}(\mathcal{D}(t)|\boldsymbol \theta) = 10^{\underline N(a_{fb}-m_0b)}\left[\prod_{i=n}^{\underline N}\dot V(t_n)f_M(m_n|b) \right]\exp[-10^{a_{fb}-bm_0}V(t)].\label{eq:partlik}
\end{equation} 
In the Appendix~\ref{sec:loglik_2}, we report the complete partial log-likelihood. Given \eqref{eq:partlik}, the posterior distribution is updated as follows:
\begin{equation}
	f''_{\boldsymbol\Theta}(\boldsymbol\theta|\mathcal{D}(t)) = c(t)\mathcal{L}(\mathcal{D}(t)|\boldsymbol \theta)f'_{\boldsymbol\Theta}(\boldsymbol\theta).\label{up_rules}
\end{equation}
The posterior $f''_{\boldsymbol\Theta}(\boldsymbol\theta|\mathcal{D}(t))$ represents the updated state of knowledge up to the time $t$ of the epistemic uncertainties. 
 
\begin{figure}[ht!]
\centering 
\includegraphics[scale=.38]{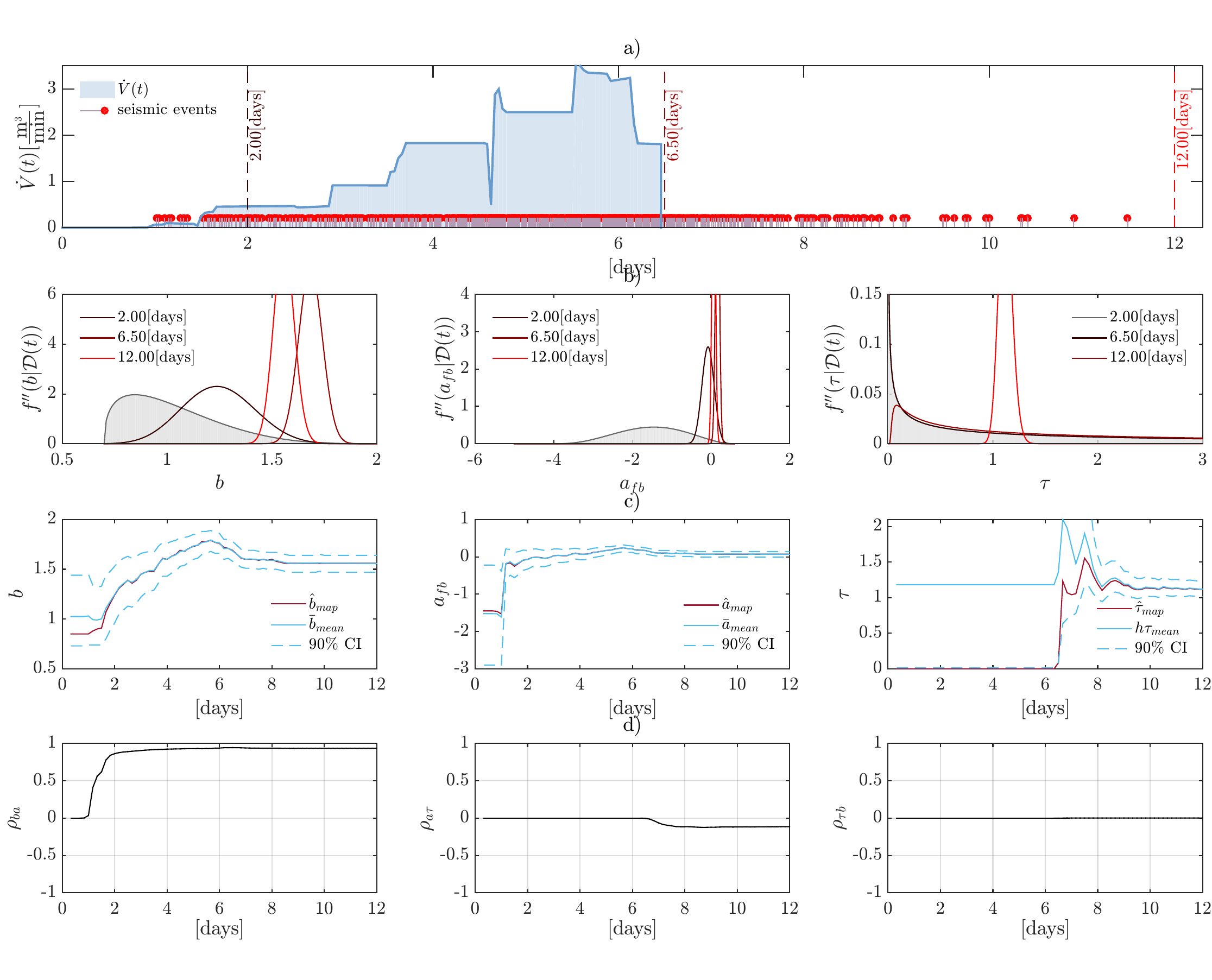}
\caption{a) Basel 2006 fluid injection profile, and seismic sequence; b) Marginal model parameter distributions: grey represents prior distributions; c) Evolution over time of the posterior mean and $map$ for each model parameter distribution; d) Evolution over time of the correlation coefficient between the model parameters}
\label{fig:epist_un}
\end{figure}

Once the operator decides to stop the injection the likelihood function reverts to its original form Eq.\eqref{eq:likelihood}, with the major exception of using the partial observation training set $\mathcal{D}(t)=[t_1,...,t_{\underline N}]$ with $t_1<...<t_{\underline N}\le t$. Once the updated likelihood has been computed, the updating rules are given by $\eqref{up_rules}$. Figure~\ref{fig:epist_un} shows the change in posterior distributions. In Animation S1, we report the full animation of the updating strategy. Based on Figure~\ref{fig:epist_un} (and Animation S1 of the Supporting Information) we can draw the following conclusions: 

\begin{itemize}
\item The epistemic uncertainties of $b$ change with time and converge to a given distribution only when injection has been terminated (Figure \ref{fig:epist_un}). The non-negligible epistemic uncertainties of $b$  {and their time evolution} should to be considered when formulating predictive equations.
\item The epistemic uncertainties of $a_{fb}$ converge rapidly towards a single value. Consequently,  {given the rate model \eqref{eq:model} and the values of $a_{fb}$ and $b$}, shortly after the injection has started, these uncertainties become negligible (Figure \ref{fig:epist_un}).
\item The epistemic uncertainties of $\tau$ change only after the injection has been terminated (Figure \ref{fig:epist_un}). Likewise for $b$, the epistemic uncertainties related to $\tau$ should not be neglected.
\item Immediately the injection starts, the joint posterior distribution captures a strong correlation between the parameters $a_{fb}$ and $b$. This is highlighted by showing the correlation coefficient as a function of time (Figure \ref{fig:epist_un}d)). Moreover, there is a weak negative correlation between $a_{fb}$ and $\tau$, while there is no correlation between $b$ and $\tau$.

\end{itemize}
 \subsection{Number and magnitude forecasting model}\label{sec:forecast}
 \subsubsection{Forecasting the number of events}
The updating rules are used also for probabilistic forecasts of the number of fluid-induced seismic events and their magnitude. Given the updated model up to time $t$ and a given, fixed timeframe $t_h= t+h$, where $h$ is a given future time window, the predictive equation for the number of events is given by: 
 \begin{equation}
P(N_{h}(t)=n_h|\mathcal{D}(t))=\int_{\boldsymbol\theta}\left[\frac{\left(\int_t^{t+h}\lambda(t'|\boldsymbol\theta)dt'\right)^{n_h}}{n_h!}\exp\left[-\int_t^{t+h}\lambda(t'|\boldsymbol\theta)dt'\right]\right]f''_{\boldsymbol\Theta}(\boldsymbol\theta|\mathcal{D}(t))d\boldsymbol\theta, \label{eq:predictive_s}
\end{equation}
where $P(N_{h}(t)=n_h|\mathcal{D}(t))=P(N_h(t'\in[t,t+h])=n_h|\mathcal{D}(t))$, with $n_h$ being the number of events in the time window $t'\in[t, t+h]$.
As shown in Section \ref{sec:BHM}, we first apply the NHPP model conditional on epistemic uncertainties; then unconditioning over $\boldsymbol \theta$ is carried out by applying the total probability theorem. Figure~\ref{fig:forecast} (and Animation S2 of the Supporting Information) show both the prediction based on $h=4$[hours] and a $90\%$ credible interval (we use 'credible' instead of 'confidence' to highlight the fact that the intervals are derived from posterior distributions), and the distribution of $P(N_{h}(t)=n_h|\mathcal{D}(t))$ for $t'\in[t,t+h]$. 
 \subsubsection{Forecasting the magnitude of the events}

\begin{figure}[ht!]
\centering 
\includegraphics[scale=.42]{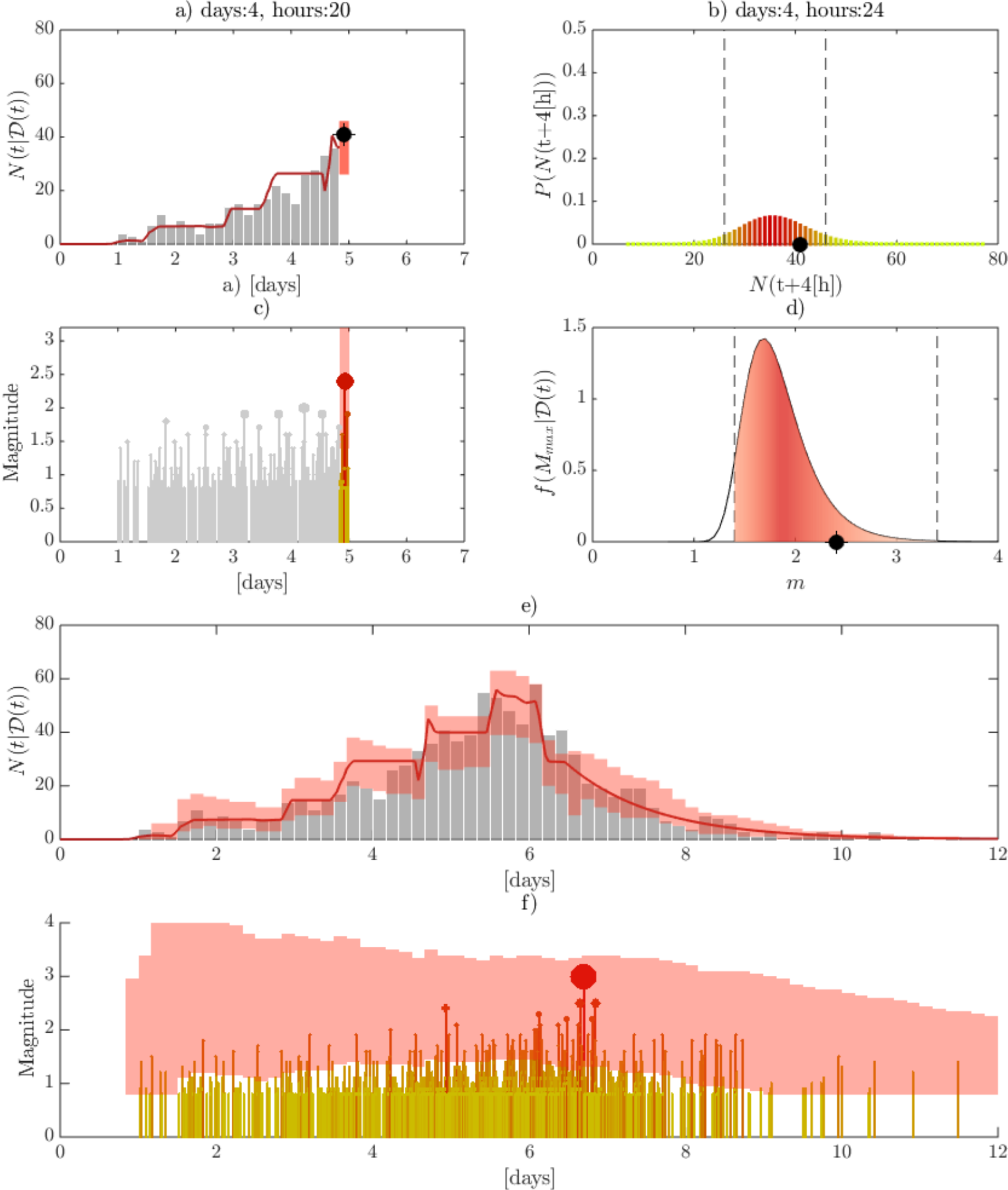}
\caption{Basel 2006 sequence: a) Prediction of the number of fluid-induced events, red bar 90\% credible interval, black dot the observed number of events; b) Distribution of the number of seismic events, grey dashed lines 90\% credible interval;  c) Time series of the magnitude events, red bar asymmetric credible interval for the $M_{max}$ in a 4[hours] time window, grey stems past seismic events, yellow-red stems observed seismic event; d) $f(m, t+h|\mathcal{D}(t))$ red area asymmetric  credible interval; e) Full prediction of the number of seismic events; d) Full prediction for $M_{max}$.}
\label{fig:forecast}
\end{figure}

  {In this subsection, } we compute the probability distribution of the maximum magnitude in the next given time windows $h$  {\citep{elst_16, Lag_16, Lag_17}}. Observe that this distribution is of particular interest in the case of a magnitude-based mitigation strategy, such as a standard traffic-light system.  {The truncated Gutenberg-Richter Magnitude distribution, \ref{eq:grd}, is used to describe magnitude frequency distribution, $f_M(m|b)$. However, in this case, $b$ is a random variable whose density changes over time, reflecting the updated state of epistemic uncertainties.} Here, the maximum magnitude in the time interval $t'\in[t,t+h]$ should not be confused with the upper limit of the Gutenberg-Richter distribution, denoted with $m_u$. Specifically, the maximum magnitude represents the random variable defined as $M_{max} = \max [M_1, M_2, ...M_{n_h},...]$ where $M_{n_h}$ is a random magnitude event in the interval $t'\in[t,t+h]$.
The distribution of $M_{max}$ can be written as follows:
\begin{equation}
P(M_{max}> m|\mathcal{D}(t)) = 1-\int_b\left[\sum_{n_h=0}^\infty(P(M>m|b)^{n_h}P(N_{t,h}=n_h|\mathcal{D}(t))\right]f''_{b}(b|\mathcal{D}(t))db\label{eq:max_distr}.
\end{equation}
In equation \eqref{eq:max_distr}, we assume that magnitude events are iid random variables. Therefore, the CCDF of $n_h$ independent events is simply $P(M>m|b)^{n_h}$.
Figure~\ref{fig:forecast} shows both the prediction and the distribution of $f(M_{max}> m|\mathcal{D}(t))$ for $t'\in[t,t+h]$ where we selected $h=4$[hours]. Since the distribution of $M_{max}$ is skewed to the right and larger magnitude events are more important, we report an asymmetric credible interval. In particular we select a $5\%$ bound for the left tail and $0.1\%$ for the right tail.

Based on Figure~\ref{fig:forecast} (and Animation S2 of the Supporting Information), we can make the following observations: 
\begin{itemize}
\item Overall, the forecast model accurately predicts the number and maximum magnitude of events in short time windows. 
\item The role of epistemic uncertainties is  important in predicting maximum magnitude distribution (Figure~\ref{fig:forecast} f)). During the initial phase the major uncertainties in $b$ (encoded in prior distribution) are reflected by large credible intervals. But as more data becomes available, the credible intervals become narrower, since the bulk of the posterior of $b$ is converging to higher values compared to the prior. However, the  credible intervals do not  {narrow} immediately after injection has terminated, but (in this case at least) only after a couple of days. This is important, because in several fluid-induced seismicity cases we observed the maximum magnitude after the shut-in event. Clearly, the proposed  {rate} model \eqref{eq:model} does not encode a physical mechanism to explain this phenomenon; so here the epistemic uncertainties of $\tau$ (which kick in once injection has stopped) are responsible for this time-delay phase. 
\item Although we did not define a decision-making criterion (which is beyond the scope of the current study), credible intervals are clearly an important tool for defining a mitigation strategy. 
\end{itemize}

\section{Conclusions}
\label{sec:conc}
In this study, we proposed a Bayesian hierarchical model for fluid-induced seismicity based on an NHPP process. We used a fluid-induced seismicity rate model proportional to the fluid injection rate and based on three physically meaningful model parameters. In conjunction with the validation of the rate model, we developed the Bayesian inference and updating rules. We showed the importance of epistemic and aleatory uncertainties and how they are encoded in the proposed framework. Finally, we developed a short-term forecasting model for the number and maximum magnitude of future fluid-induced seismic events over a given, fixed time window. The results showed that the Bayesian hierarchical model is a suitable and robust probabilistic framework for classifying, analyzing, and forecasting the uncertainties related to fluid-induced seismicity. An interesting extension of the study would involve investigating  the use of rate models based on geomechanical principles. Here, the outlined framework would serve as a powerful tool for analyzing the importance of uncertainties associated with physical-model parameters and for uncovering their correlation structure. 

\break
\newpage



%
%


%
%
%

\appendix
\section{Appendix}
\subsection{Complete log-likelihood function}\label{sec:loglik} 
Given~\eqref{eq:likelihood}, and assuming a  truncate Gutenber-Richter Magnitude frequency distribution,
\begin{linenomath*}
\begin{equation}\label{eq:grd}
f_M(m|b) = b\ln(10)\frac{10^{-b(m-m_{0})}}{(1-10^{-b(m_{u}-m_{0})})},\text{ for $m_0\le m \le m_u$}
\end{equation}
\end{linenomath*}
where $m_u$ is the upper bound of the magnitude distribution, the complete log-likelihood can be written as follow
\begin{linenomath*}
\begin{equation}
	\begin{split}
		\ln\mathcal{L}(\mathcal{D}| \boldsymbol \theta)&=\sum_{n=1}^{N}\ln \lambda(t_n|\boldsymbol\theta)+\sum_{n=1}^N\ln f_M(m_n|b)-\Lambda(T|\boldsymbol\theta)\\
		&= \frac{ N(a_{fb}-bm_0)}{\log e}+\sum_{n=1}^{\underline N}\ln \dot V(t_n)+(N-\underline N)\ln\dot V(t_s)-\sum_{\underline N+1}^N
		\left(\frac{t_n-t_s}{\tau}\right )-\\  & 10^{a_{fb}-bm_0}\left[V(t_s)+\dot V(t_s)\tau\left(1-\exp\left(-\frac{T-t_s}{\tau}\right)\right)\right]+N\ln(b)+N\ln(\ln(10))\\
		&-b\ln(10)\sum_{n=1}^Nm_n- N\ln(10^{-bM_0}-10^{-bM_{max}})
    \end{split}
\end{equation} 
 \end{linenomath*}
 where $\underline N$ is the number of seismic event before the injection is terminated. 
 \subsection{Partial log-likelihood function}\label{sec:loglik_2} 
 The partial log-likelihood before the shut-in event can be written as follow
\begin{equation}
	\begin{split}
		\ln\mathcal{L}(\mathcal{D}(t)|  \boldsymbol \theta)&=\sum_{n=1}^{\underline{N}}\ln \lambda(t_n|\boldsymbol\theta)+\sum_{n=1}^{\underline N}\ln f_M(m_n|b)-\Lambda(T|\boldsymbol\theta)\\
		&= \frac{ \underline{N}(a_{fb}-bm_0)}{\log e}+\sum_{n=1}^{\underline N}\ln \dot V(t_n)-10^{a_{fb}-bm_0} V(t)+\underline N\ln(b)+\underline N\ln(\ln(10))\\
		&-b\ln(10)\sum_{n=1}^{\underline N}m_n- \underline N\ln(10^{-bM_0}-10^{-bM_{max}})
    \end{split}
\end{equation}

%
%

%


%

%

\acknowledgments

We gratefully acknowledge the Swiss Competence Center for Energy Research Supply of Electricity (SCCER--SoE) for supporting this research. Moreover, this work was supported by DESTRESS. DESTRESS has received funding from the European Union's Horizon 2020 research and innovation programme under grant agreement No.691728.
 The seismic catalogs used in this study have different sources and are all public available. The sources are listed in Table 1 of the Supporting Information. Finally, we thank the anonymous reviewers for their assistance in evaluating this paper.

%
%
%
%
%
%
%
%
%




\listofchanges

\end{document}